# REPRODUCIBILITY, ENERGY EFFICIENCY AND PERFORMANCE OF PSEUDORANDOM NUMBER GENERATORS IN MACHINE LEARNING: A COMPARATIVE STUDY OF PYTHON, NUMPY, TENSORFLOW, AND PYTORCH IMPLEMENTATIONS

Benjamin ANTUNES

David R.C. Hill

LIMOS, CNRS, Clermont Auvergne INP Université
Clermont Auvergne
Mines St-Etienne
benjamin.antunes@uca.fr

LIMOS, CNRS, Clermont Auvergne INP Université
Clermont Auvergne
Mines St-Etienne
david.hill@uca.fr

## ABSTRACT

Pseudo-Random Number Generators (PRNGs) have become ubiquitous in machine learning technologies because they are interesting for numerous methods. The field of machine learning holds the potential for substantial advancements across various domains, as exemplified by recent breakthroughs in Large Language Models (LLMs). However, despite the growing interest, persistent concerns include issues related to reproducibility and energy consumption. Reproducibility is crucial for robust scientific inquiry and explainability, while energy efficiency underscores the imperative to conserve finite global resources. This study delves into the investigation of whether the leading Pseudo-Random Number Generators (PRNGs) employed in machine learning languages, libraries, and frameworks uphold statistical quality and numerical reproducibility when compared to the original C implementation of the respective PRNG algorithms. Additionally, we aim to evaluate the time efficiency and energy consumption of various implementations. Our experiments encompass Python, NumPy, TensorFlow, and PyTorch, utilizing the Mersenne Twister, PCG, and Philox algorithms. Remarkably, we verified that the temporal performance of machine learning technologies closely aligns with that of C-based implementations, with instances of achieving even superior performances. On the other hand, it is noteworthy that ML technologies consumed only 10% more energy than their C-implementation counterparts. However, while statistical quality was found to be comparable, achieving numerical reproducibility across different platforms for identical seeds and algorithms was not achieved.

## 1    INTRODUCTION

Contemporary machine learning (ML) researchers predominantly use high-level programming languages and frameworks to conduct their studies. Python is the principal programming language in ML, leading to the widespread adoption of frameworks such as PyTorch and TensorFlow, often coupled with NumPy. In this paper, we want to study the statistical quality, reproducibility, energy and time consumption of the pseudo random number generation in these technologies. The literature on the quality of Pseudo-random number generators (PRNGs) within ML technologies remains sparse; our investigation addresses this gap.

In Python, the default PRNG algorithm used is Mersenne Twister (MT) (Matsumoto & Nishimura, 1998). In TensorFlow, the default PRNG algorithm is Philox (Threefry from the same family of crypto secure generator is also available) (Salmon et al., 2011), similarly to PyTorch. NumPy offers a variety of PRNGs, and thus more flexibility. The default PRNG algorithm proposed by NumPy is PCG (O'neill, 2014). For our study, we check and compare reproducibility, performance, statistical quality and energy consumption, for the following PRNGs: MT, Philox, PCG and Mrg32k3a (L'ecuyer, 1999) as a reference. We use the original C implementations provided by the PRNGs authors.

As described in (Antunes & Hill, 2023), Salmon et al. introduced the Philox, Threefry and ARS algorithms at the 2011 Supercomputing Conference, they incorporate cryptographic techniques akin to AES (Daemen & Rijmen, 2000). Although their cryptographic nature makes them relatively slow, their statistical properties are commendable, albeit with some reproducibility issues (Li, 2012). MRG32k3a, devised by L'Ecuyer in 1999, is a combined recursive pseudo-random number generator chosen specifically since it was built to obtain the best statistical results when faced to TestU01, the most complete statistical test battery developed to assess PRNGs (L'ecuyer & Simard, 2007). This software proposes more than 100 tests at the "big Crush" level, it will be discussed below. MRG32k3a can be significantly slower than the famous Mersenne Twister, 15 to 20 times slower when comparing optimized C implementations. PCG, developed in 2014 by O'Neill, is touted for its superior statistical attributes compared to other generators, this could not be confirmed with a thorough TestU01 campaign. The initial Mersenne Twister generator was introduced in 1998 by Matsumoto and Nishimura, it has known limits but is renowned for its long period. Its 2002 iteration improved its initialization. SFMT version, designed by Saito & Matsumoto in 2006, capitalizes on modern processor capabilities and offers twice more speed and even superior statistical qualities. A GP-GPU version was proposed and is known as MTGP. However, it is important to note that the Mersenne Twister family is not apt for cryptographic applications. Thought it presents some minor statistical flaws, we are not aware of applications that have been impacted and it is particularly well spread in many scientific libraries.

To assess the quality of a Pseudorandom Number Generator (PRNG), statistical evaluations are employed to distinguish between superior and inferior PRNGs. Historically, Donald Knuth introduced an initial array of statistical evaluations for PRNGs in the second volume of "The Art of Computer Programming". Despite their age, these tests remain relevant both for academic pursuits and as a reference (Knuth, 1973). In 1996, Marsaglia introduced a concise suite comprising 15 tests known as Die Hard. The original source code for Die Hard is no longer available. However, a reference to the historical code can be found via a "wayback machine" link provided in the references section (Marsaglia, 1996). Brown, along with his Australian associates, extended Marsaglia's work and introduced an updated set of tests, released as open source software. This suite was aptly named Die Harder (Brown et al., 2013). The National Institute of Standards and Technology's (NIST) Statistical Test Suite is regarded as the benchmark for assessing random and pseudorandom number generators, especially in cryptographic contexts (Rukhin et al., 2001). L'Ecuyer and Simard unveiled an open source library dedicated to the empirical evaluation of random number generators. Known as TestU01 as previously cited, this suite offers a comprehensive array of tests, categorized into various levels of scrutiny such as Small Crush, Crush, and Big Crush, among others (L'Ecuyer & Simard, 2007). To measure the quality of pseudo random numbers generated in ML technologies, we used the Big Crush test battery from TestU01, consisting in 106 statistical tests. Random sampling is particularly interesting in training artificial intelligence models. In the category of "General Game Playing", where machines must play a new game starting with its basic rules, an annual competition is organized by Stanford. In this field, the evolution of machine capabilities has allowed the stochastic approach of Monte Carlo Tree Search (MCTS) to become more and more efficient. In particular, as of 2007, these methods have proven to be very successful in the game of Go, and it is interesting to note that all world champion programs in two-player GGP now use MCTS (Cazenave & Jouandeau, 2009).

The rise of deep learning and complex models in ML necessitates efficient computational resources to process vast amounts of data. Hardware accelerator manufacturers are racing to propose better performances at an impressive pace. Performance, often quantified by the time taken to compute or the speed of operations, directly impacts the feasibility of training larger models and iterating over them during the research phase. While an optimized algorithm or efficient hardware can improve time efficiency, the energy consumed during computations also becomes a significant concern, especially with the current emphasis on environmental sustainability (Goralski & Tan, 2020). High energy consumption not only leads to higher operational costs but also contributes to increased carbon footprints in data centers. Therefore, understanding and optimizing the performance and energy efficiency of computations, including those of PRNGs, are imperative. Efficient PRNGs can lead to faster initializations, shuffling, and other stochastic operations in ML workflows, further reducing both time and energy consumption.

Another aspect of science advancement has to be tackled: reproducibility as a cornerstone of scientific integrity (Hill, 2015). It enables researchers to validate, build upon, or challenge prior findings. In the realm of ML, reproducibility ensures that results obtained in one run can be consistently achieved in subsequent runs, given the same configurations. This consistent outcome is crucial for debugging, model comparison, validation, and ensuring the reliability of the technology in real world applications. PRNGs play a pivotal role in this context. Since many ML processes, from data splitting to weight initialization, rely on pseudorandom sequences, the reproducibility of PRNG outputs is vital. Without repeatable and consistent PRNG outputs, subtle differences can amplify through the training process, leading to markedly different outcomes. Beyond individual experiments, reproducibility is also vital for the broader scientific community (Nature Editorial, 2024). When results can be reliably reproduced, it fortifies the foundation upon which future research is built, ensuring a progressive and trustworthy scientific trajectory.

This paper aims to answer to the following questions:

- Are PRNGs implemented in ML frameworks giving the same results as their initial C codes proposed by the original PRNGs implementations when identically initialized?
- Does pseudo-random numbers generated with ML main language, libraries and frameworks have the same statistical quality than those produced by the original code given by the PRNG authors?
- Is the process of generating random numbers in ML frameworks more time-consuming when compared to the original C codes?
- Does random number generation within ML frameworks require more energy than its C code counterparts?
- Taking into account the previous points, is there a consistency between the performance of 32-bit integer and 64-bit double precision of the generated numbers?

Our discussion will begin with an overview of prior research on the application of stochastic processes in machine learning. Subsequently, we will present the method employed in our experiments. Following this, we will present the findings about time performance, energy consumption by minutes, overall energy consumption and numerical reproducibility. Finally, we talk about on the implications and future directions of our results.

## 2    THE IMPORTANCE OF PRNGS IN MACHINE LEARNING

To underline the importance of the PRNG statistical quality on the neural network training, a recent work from Huk (Huk et al., 2021) attempted to quantify the potential differences in classification performance of CNNs and MLPs when varying the PRNG. They draw the 95% confidence interval for each quality measurement, for different PRNGs. The results indicated minor variations in quality associated with different PRNGs, as evidenced by non-overlapping confidence intervals. This study shows that the PRNG algorithm used might have an incidence (needing to double the confidence intervals of evaluation metrics) over the quality of the neural network training. (Koivu et al., 2022) also shows a correlation between the statistical quality of a PRNG and the resulting quality of the dropout method applied to the neural network. Additional research is necessary to explore various neural network architectures, and assess the impact of PRNG quality on neural network performance, and replicate these results, given the scarcity of literature on this topic. The quality of the PRNGs used in ML is not well studied, and it would be interesting to investigate. Indeed, stochastic processes has become increasingly important in ML over the years due to its efficiency in some cases. As a result, PRNGs have become indispensable in ML technologies.

To illustrate the importance of PRNGs in machine learning, we consider multiple stochastic methods such as the Stochastic Gradient Descent (SGD). It is a cornerstone optimization algorithm for training models in machine learning and deep learning. It operates by using a single or a small batch of training samples to calculate the gradient and update parameters, rather than using the entire training dataset. Bottou demonstrated the versatility of SGD across various ML methodologies, including perceptrons and multi-layer networks (Bottou, 2003). Bottou (Bottou, 2010) highlighted the role of SGD role in addressing the constraints of massive data volumes and limited computational power, while Nguyen et al. (Nguyen et al., 2017) introduced "SARAH" (StochAstic Recursive grAdient algoritHm) to refine the initial SGD.

Beyond the commonly employed SGD algorithm, known for its efficiency, it's worth noting the significant role of regularization techniques that have demonstrated considerable utility and similarly require elements of randomness. Dropout is one such regularization strategy tailored for neural networks to mitigate overfitting. Overfitting transpires when a model excessively conforms to training data, compromising its ability to generalize, which results in subpar performance on novel data. Dropout addresses this by randomly omitting a selection of neurons and their connections throughout the training process (Srivastava et al., 2014).

Additionally, the concept of stochastic depth, another regularization technique reliant on randomness, was designed to overcome obstacles inherent in training deep convolutional networks, such as vanishing gradients and protracted training durations. It streamlines the training process by randomly omitting a set of layers in each training batch and seamlessly connecting the remaining ones using the identity function, thus reducing training time and potentially increasing test accuracy (Huang et al., 2016).

Randomness is also instrumental in data augmentation, a method aimed at expanding the data set by incorporating modified replicas of existing data or generating new synthetic data. This approach is particularly beneficial in machine learning, enhancing model performance through a more robust dataset. For image-related tasks, data augmentation can involve alterations like rotation, cropping, or flipping. Notable algorithms that employ data augmentation include the Expectation-Maximization algorithm, the algorithm for posterior sampling, and Markov chain Monte Carlo methods for posterior sampling (Van Dyk & Meng, 2001). In deep learning for images, augmentation techniques that incorporate randomness span a wide spectrum, from geometric adjustments and color space alterations to kernel filters, image mixing, random erasing, and even neural style transfer. Moreover, test-time augmentation introduces variability during model evaluation, which is critical for enriching datasets and fortifying model resilience (Shorten & Khoshgoftaar, 2019).

Additionally, the concept of bootstrapping complements these techniques by providing another layer of randomness and robustness. Bootstrapping, involving the creation of multiple subsets of the dataset through sampling with replacement, allows for the generation of diverse training conditions. This technique is instrumental in enhancing model accuracy and stability, particularly in ensemble learning methods where it contributes to a more comprehensive exploration of the data space and better generalization of the model (Tsamardinos et al., 2018).

A recent survey highlights the pervasive application of randomness in machine learning as a trade-off for hardware efficiency and computational performance (Liu et al., 2020). The usage of PRNGs in machine learning is wildly spread. Examples include Bayesian neural networks (Neal, 2012), Variational autoencoders presented in (Kingma & Welling, 2013) and Reinforcement Learning (Sutton & Barto, 2018). Additionally, some methods propose the injection of gradient noise as a strategy to enhance deep neural network training (Neelakantan et al., 2015).

Some recent work are more focused on the use of pseudo random generation and the power consumption of neural networks. In (Kim et al., 2016), they used stochastic computing on deep neural networks and obtained better results for latency and power consumption. In this case, the old stochastic computing (SC) approach, originally introduced by John Von Neumann in the beginning of the sixties, where information is represented and processed using random bit streams, serve for complex computations operated with bit-wise operations. In (Liu et al., 2018), authors point out that SC can be costly in term of energy efficiency when used in deep neural networks.

Furthermore, the evolving landscape of machine learning has seen the rise of Transformer architectures (Vaswani et al., 2017), especially in the domain of large language models. These architectures, exemplified by models like GPT (Generative Pre-trained Transformer), still rely on randomness in their training phase.

This randomness manifests in the form of stochastic gradient descent and dropout techniques, essential for preventing overfitting and promoting model generalization.

With this short literature review, we can confirm that randomness, along with PRNGs, are prominent artificial intelligence technologies that will become ubiquitous in our lives. Since the quality of pseudo-random numbers in machine learning frameworks remains under-explored, as our literature search yielded no relevant studies, we want to bridge this knowledge gap.

## 3    MATERIALS AND METHODS

To address the questions raised in introduction, we selected prominent ML frameworks, specifically PyTorch and TensorFlow, along with the Python and the NumPy library due to their widespread use in the ML field. For benchmark purposes, we have retained the original C code implementations of Mersenne Twister, PCG, Philox, and Mrg32k3a as a standard of comparison (all codes are proposed on the authors' web pages).

The Mersenne Twister supports native generation of both 32-bit integers and 64-bit doubles. On the other hand, Mrg32k3a is limited to generating only 64-bit doubles. In order to maintain fidelity to the original implementations, we restricted our use of Mrg32k3a to experiments involving 64-bit doubles. Conversely, the Philox algorithm was only available for generating 32-bit integers from its authors. PCG offers the possibility for both, but the author prefers to stick with integer "*Like the Unix rand and random facilites, this library does not provide a direct facility to generate floating point random numbers. It turns out that generating random floating point values is surprisingly challenging.*" ( https://www.pcg-random.org/using-pcg-c-basic.html ). However, as the author provide a solution to generate double, we used PCG in both cases, like MT. ML frameworks, with their advanced APIs, allow for the straightforward generation of either 32-bit integers or 64-bit doubles. The most recent version of TensorFlow suggests using a Generator object, which we explicitly applied to the Philox algorithm. For PyTorch, while the underlying algorithm is believed to be Philox based on documentation, the user cannot specify his generator choice. NumPy stands out as perhaps the most versatile library for handling various PRNGs, offering clear documentation and a range of available algorithms. With NumPy we used the Generator object, setting it to explicitly use Mersenne Twister, Philox, and PCG.

These technologies differ from traditional scientific computing practices in C, C++, or Fortran, where random numbers are typically generated individually as needed. In contrast, ML frameworks are optimized to generate random numbers in bulk as part of tensor objects (akin to matrices). Therefore, we conducted experiments both ways: generating numbers one by one and in bulk. For Python, the most efficient approach was to generate numbers individually.

As PCG propose different versions, for 64 bits we choose the exact same version as NumPy (PCG 128/64 XSL-RR) and for 32 bits we used PCG 64/32 XSH-RR.

We initialized all PRNGs with the same seed value. To neutralize language-specific data type disparities, we used the seed value '0', ensuring a zero-filled seed memory pointer across different data types. Although initializing with zero can be problematic for some PRNGs (Saito & Matsumoto, 2006), this was intentionally done to observe the resultant behavior. It is imperative for researchers in the scientific community to recognize that a seed and the complete state of a PRNG are distinct entities. The state of the PRNG is determining the output value it generates. In contrast, utilizing a seed involves the application of a specific function to convert the seed into the full state of the PRNG. It is noteworthy that this transformation process may vary across different technological platforms. Given that the entire machine learning framework is fundamentally dependent on the seeding function, our study is primarily focused on studying this aspect.

Our evaluation utilized various Bash scripts: one to run time and energy consumption assessments—generating $2^{30}$ numbers one by one or at once and timing the process with the 'time' command. Energy consumption was monitored over a set period (e.g., 30 seconds), with results extrapolated over the entire duration. We replicated these measurements 30 times to strengthen the statistical validity of our measures, this leads to the study of samples of a bit less than $2^{35}$ numbers.

Energy measurements were obtained using PowerJoular (Noureddine, 2022). This tool offers the possibility to measure the energy consumption of a given Process ID, using RAPL Intel feature (David et al., 2010), also available on recent AMD chips. We compiled all C codes with different optimization levels (none, -O2, and -O3) to discern the impact of compiler optimizations on time and energy efficiency.

For quality evaluation, we ran another set of Bash scripts designed for extensive testing. The TestU01 BigCrush test battery, which typically requires a bit more than $2^{38}$ numbers based on TestU01 documentation, prompting us to generate $2^{39}$ numbers (one order of magnitude over). Given that BigCrush is not designed to read numbers from a file in its original form, we made a C-code interface. We stored the ML-generated numbers in a binary file and subsequently, the C program reads the numbers sequentially from this file to provide the inputs required by BigCrush. This method was also applied to the PRNGs coded in C for a fair comparison. Preliminary tests showed no significant difference between the modified approach and the original, confirming the validity of our method. However, it is important to note that storing $2^{39}$ doubles take 4.4TB of storage and 2.2TB for 32 bits integers.

Finally, for numerical reproducibility, we generated 100 pseudo random numbers in a readable file, and computed "diff" command over files, the algorithm being the same, seeded identically, we expect bitwise identical results.

We used a Jupyter Notebook to analyze all the results, and run all bash scripts to easily reproduce the experiments. All codes and data are available at: https://gitlab.isima.fr/beantunes/random-numbers-in-machine-learning/

Experiments were performed on a machine with two AMD 7763 64-cores processors, leading to 128 physical cores and 256 logical cores. The machine has 512GB of RAM, and 7.7TB of NVMe storage. We had root access, so we were able to perform energy consumption measurements (RAPL needs root access to be used). The Python version used is 3.11.5. The GCC version used is 13.2.0. The operating system is a Linux, Debian 6.4.13-1.

## 4    RESULTS

### 4.1    Time performance

Tables 1 and 2 illustrate the time required to generate $2^{30}$ numbers in each experiment. First, distinct performance discrepancies between 32-bit integers and 64-bit doubles are observed. Notably, the PCG algorithm demonstrates superior speed for 32-bit integers but requires to quadruple its generation time for 64-bit doubles. The Mersenne Twister code, in its original implementation, takes the same time for both. When implemented using NumPy, the MT algorithm demonstrates a pronounced divergence in generation time, taking approximately 4.5 seconds for 32-bit integers versus 13 seconds for 64-bit doubles (for 1 billon drawings), whereas the original version maintains a consistent 4-second duration for each. However, we can see that PRNG implementations via ML Python frameworks have a good computational efficiency, as Python and C code execution times are mixing in the performance rankings. However, the MT algorithm is significantly slower in pure Python. For the PCG and Philox algorithms, implementations utilizing ML technologies appear to outperform the original versions (in C code), despite the use of –O2 or –O3 compilation optimizations (when we were able to use them, because sometimes, the usage of compilation optimization leads to the malfunction of the code). The primary distinction between the original C code and ML-based code lies in the unsuitability of the latter for generating numbers sequentially, resulting in significantly poor performance when trying to generate the random numbers sequentially and, in the case of TensorFlow, an infeasibility due to RAM overload, despite the availability of more than 500GB of RAM in our test.

| Generator | Real time (s) | Real time 95% CI | User time (s) | User time 95% CI |
|---|---|---|---|---|
| pcg32Integer | 2,45 | [2,27; 2,64] | 2,45 | [2,27; 2,63] |
| numpyIntegerTasksetAtOnce | 2,60 | [2,59; 2,60] | 2,20 | [2,19; 2,22] |
| tensorflowIntegerAtOnce | 3,22 | [3,19; 3,25] | 17,89 | [17,70; 18,08] |
| numpyIntegerAtOnce | 3,42 | [3,23; 3,61] | 3,98 | [3,80; 4,15] |
| mt19937arIntegerO3 | 4,29 | [4,17; 4,42] | 4,29 | [4,17; 4,41] |
| numpyIntegerMtAtOnce | 4,55 | [4,42; 4,68] | 5,15 | [5,02; 5,27] |
| mt19937arIntegerO2 | 4,74 | [4,68; 4,81] | 4,74 | [4,67; 4,81] |
| numpyIntegerPhiloxAtOnce | 6,77 | [6,63; 6,92] | 7,37 | [7,23; 7,51] |
| tensorflowIntegerTasksetAtOnce | 7,08 | [7,04; 7,13] | 6,23 | [6,20; 6,27] |
| mt19937arInteger | 7,10 | [6,96; 7,24] | 7,10 | [6,95; 7,24] |
| pytorchIntegerTasksetAtOnce | 8,06 | [8,00; 8,13] | 7,12 | [7,06; 7,18] |
| pytorchIntegerAtOnce | 9,09 | [8,99; 9,19] | 8,93 | [8,85; 9,01] |
| philoxInteger | 90,06 | [89,74; 90,39] | 90,06 | [89,73; 90,38] |
| pythonIntegerOneByOne | 425,92 | [424,24; 427,60] | 425,91 | [424,23; 427,58] |
| pythonIntegerTasksetAtOnce | 486,11 | [484,14; 488,09] | 452,94 | [451,32; 454,55] |
| pythonIntegerAtOnce | 489,02 | [487,30; 490,75] | 453,29 | [451,88; 454,70] |
| pytorchIntegerOneByOne | 2281,79 | [2248,98; 2314,61] | 2282,33 | [2249,51; 2315,16] |
| numpyIntegerMtOneByOne | 6327,76 | [6228,26; 6427,27] | 6328,50 | [6228,97; 6428,02] |
| numpyIntegerOneByOne | 6458,61 | [6396,91; 6520,31] | 6459,50 | [6397,77; 6521,23] |
| numpyIntegerPhiloxOneByOne | 6552,21 | [6472,50; 6631,91] | 6553,13 | [6473,41; 6632,85] |

Table 1: Real time and user time taken for each experiments, for $2^{30}$ 32 bits integer random number generation

Finally, when comparing User time and Real time, we can see that TensorFlow is the only technology that is using implicit parallelization. We can then suppose that if less cores were available in the machine, or if the machine was overloaded due to some other running processes, the TensorFlow generation would have taken more time than NumPy and similar to PyTorch, due to the fact that doing parallelization on an already overloaded machine will not improve performance, and can even worsen them. Using taskset to set affinity of a single process to only one core shows a slight improvement for ML frameworks, except for TensorFlow, due to its native implicit parallelization.

In tables 3 and 4, our observations also extended to the time required to generate and store $2^{39}$ pseudorandom numbers (in minutes). As anticipated, the duration for generating and storing these numbers is approximately half as long for 32-bit values compared to 64-bit values. Notably, the Mrg32k3a generator exhibits the slowest performance over C-coded generators, although it successfully passes all statistical benchmarks. We can notice that the PCG generator is faster than Mrg32k3a, and sometimes "Crush resistant". It is unexpectedly clear that generating integers with Python is considerably more time-consuming; in both 32-bit and 64-bit instances, it is the least efficient technology (using MT algorithm). For the creation of $2^{39}$ numbers, we employed a strategy that favors ML frameworks inclined towards blocking: generating segments of $2^{20}$ numbers sequentially until the full $2^{39}$ was reached. Among these frameworks, TensorFlow demands the most system and user time, a likely consequence of its underlying parallelization which could be problematic on limited computational resources. Interestingly, ML frameworks demonstrate competitive performance relative to C implementations. This outcome was unforeseen and underscores the high degree of optimization present in these advanced-level frameworks.

| Generator | Real time (s) | Real time 95% CI | User time (s) | User time 95% CI |
|---|---|---|---|---|
| tensorflowAtOnce | 3,38 | [3,28; 3,47] | 32,33 | [31,90; 32,76] |
| mt19937arO3 | 4,20 | [4,06; 4,34] | 4,20 | [4,06; 4,34] |
| numpyTasksetAtOnce | 4,35 | [4,32; 4,39] | 3,56 | [3,52; 3,61] |
| mt19937arO2 | 4,50 | [4,34; 4,67] | 4,50 | [4,34; 4,67] |
| well19937O3 | 4,96 | [4,85; 5,08] | 4,96 | [4,85; 5,07] |
| well19937O2 | 4,97 | [4,83; 5,12] | 4,97 | [4,83; 5,12] |
| numpyAtOnce | 5,77 | [4,73; 6,82] | 5,75 | [4,71; 6,79] |
| pytorchTasksetAtOnce | 6,02 | [5,94; 6,11] | 5,31 | [5,25; 5,36] |
| pytorchAtOnce | 6,90 | [6,76; 7,05] | 7,01 | [6,90; 7,12] |
| mt19937ar | 7,48 | [7,31; 7,66] | 7,48 | [7,31; 7,66] |
| tensorflowAtOnce | 8,18 | [8,12; 8,24] | 6,67 | [6,63; 6,72] |
| pcg64O3 | 11,00 | [10,83; 11,17] | 11,00 | [10,83; 11,16] |
| pcg64O2 | 11,07 | [10,92; 11,23] | 11,07 | [10,91; 11,23] |
| numpyMtAtOnce | 13,08 | [11,56; 14,60] | 7,27 | [7,16; 7,38] |
| well19937a | 13,08 | [13,08; 13,09] | 13,08 | [13,07; 13,09] |
| pcg64 | 13,18 | [13,05; 13,31] | 13,18 | [13,05; 13,31] |
| numpyPhiloxAtOnce | 13,26 | [12,59; 13,92] | 12,00 | [11,89; 12,11] |
| mrg32k3aO3 | 19,97 | [19,80; 20,14] | 19,96 | [19,79; 20,13] |
| mrg32k3aO2 | 31,47 | [31,21; 31,72] | 31,46 | [31,21; 31,71] |
| pythonOneByOne | 36,87 | [36,23; 37,51] | 36,86 | [36,22; 37,50] |
| mrg32k3a | 43,13 | [42,96; 43,29] | 43,12 | [42,96; 43,29] |
| pythonTasksetAtOnce | 69,52 | [69,02; 70,03] | 43,89 | [43,66; 44,12] |
| pythonAtOnce | 75,46 | [72,06; 78,86] | 48,48 | [47,23; 49,73] |
| numpyMtOneByOne | 319,89 | [318,05; 321,74] | 320,83 | [318,99; 322,67] |
| numpyPhiloxOneByOne | 323,06 | [321,28; 324,85] | 323,98 | [322,20; 325,76] |
| numpyOneByOne | 330,98 | [326,41; 335,54] | 331,88 | [327,31; 336,44] |
| pytorchOneByOne | 2388,41 | [2381,38; 2395,44] | 2388,85 | [2381,80; 2395,90] |

Table 2: Real time and user time taken for each experiments, for $2^{30}$ 64 bits double random number generation

| File | Real Time | User Time | Sys Time |
|------|-----------|-----------|----------|
| timeIntegerNumpySaving | 91,2 | 27,6 | 63,4 |
| timeIntegerNumpyMtSaving | 97,6 | 36,4 | 61,1 |
| timeIntegerPytorchSaving | 121 | 41,6 | 79,4 |
| timeIntegerNumpyPhiloxSaving | 125,6 | 57,6 | 67,8 |
| timeIntegerTensorflowSaving | 131,4 | 826,9 | 149,6 |
| timeIntegerPcgSaving | 164,4 | 98,1 | 66 |
| timeIntegerMtSaving | 180,9 | 107,2 | 73,4 |
| timeIntegerPhiloxSaving | 229,8 | 150,8 | 78,8 |
| timeIntegerPythonSaving | 4957,1 | 4890,9 | 65,5 |

Table 3: Time taken to save $2^{39}$ 32 bits integer numbers for each frameworks.

| File | Real Time | User Time | Sys Time |
|------|-----------|-----------|----------|
| timeNumpySaving | 170,6 | 33,1 | 137,3 |
| timePytorchSaving | 176,2 | 43,2 | 132,5 |
| timeNumpyMtSaving | 177,5 | 48,4 | 128,6 |
| timeNumpyPhiloxSaving | 231,9 | 96,7 | 135,0 |
| timeMtSaving | 281,8 | 126,0 | 154,4 |
| timeWellSaving | 283,0 | 127,4 | 155,2 |
| timeTensorflowSaving | 288,8 | 1893,2 | 393,4 |
| timePcgSaving | 346,8 | 185,8 | 160,6 |
| timeMrgSaving | 449,9 | 307,2 | 142,2 |
| timePythonSaving | 1355,8 | 1218,5 | 136,7 |

Table 4: Time taken to save $2^{39}$ 64 bits double numbers for each frameworks.

## 4.2 Energy consumption by minutes

In tables 5 and 6, the energy consumption is presented in terms of Joule by minutes for each experiment, corresponding to 32-bit integers and 64-bit doubles, respectively. From these findings, it is obvious that ML technologies consume around 10% more energy than traditional C code implementations. We notice that the PRNG Philox is identified as a particularly high-energy-consuming algorithm relative to its counterparts. It is supposed to be crypto-secure, a characteristic typically associated with an increased computing time because of additional complexity. However crypto-secure pseudo-random number generators (CS-PRNGs), even though found traditionally slower, have a high rate of success in passing statistical tests. According to the paper by M. O'Neill on PCG, the variants employed in this study (PCG 128/64 XSL-RR and PCG 64/32 XSH-RR) are also claimed to be crypto-secure. They are supposed to be successful to pass the big crush TestU01 battery but we encountered problems. Further investigation would be needed to ensure the claim the PCG is crypto-secure, for instance with the NIST Statistical Test Suite (STS) but this is out of the scope of our paper. We can also see that, for all ML frameworks, the block generation "at once" uses less energy than generating "one by one", especially in the case of 32 bits integers. In the next section, we will talk about the energy consumption, but based on the real time taken to compute, not by minutes. Figure 1 outlines the differences between the energy consumption by minutes between the different versions (C code and ML library or framework) of an implemented PRNG algorithm. Overall, we compare the energy consumption of all C implementations and all Python implementations, resulting in around 20% more energy consumed by minutes by Python implementations. In this figure, we did not take into account one by one numbers generation from NumPy because of they consume much more energy per minute (see table 5 and 6). This difference is probably due to the low efficiency of NumPy to generate one by one pseudo random numbers, the blocking approach being much more efficient.

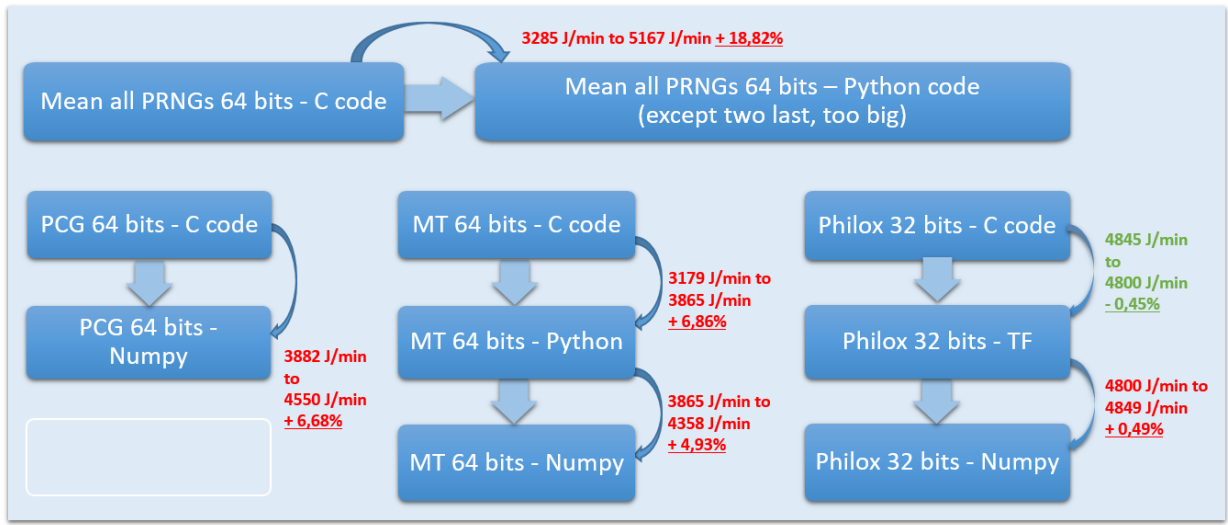

Figure 1: Consumption by minute in Joule - Differences between C code implementations and Python implementations, for each PRNG.

| Generator | Energy consumption (J/min) | Energy consumption 95% CI |
|---|---|---|
| pcg32Integer | 3209,70 | [3185.35; 3234.05] |
| mt19937arIntegerO2 | 3323,89 | [3226.38; 3421.40] |
| mt19937arInteger | 3419,13 | [3285.98; 3552.27] |
| mt19937arIntegerO3 | 3607,74 | [3476.79; 3738.68] |
| pythonIntegerTasksetAtOnce | 4298,72 | [4260.00; 4337.43] |
| pythonIntegerAtOnce | 4375,21 | [4110.53; 4639.88] |
| numpyIntegerAtOnce | 4688,14 | [4669.67; 4706.61] |
| numpyIntegerTasksetAtOnce | 4766,35 | [4747.59; 4785.10] |
| pytorchIntegerTasksetAtOnce | 4766,96 | [4748.06; 4785.86] |
| tensorflowIntegerTasksetAtOnce | 4800,25 | [4784.14; 4816.36] |
| pytorchIntegerAtOnce | 4812,53 | [4775.33; 4849.73] |
| philoxInteger | 4845,48 | [4824.15; 4866.81] |
| numpyIntegerMtAtOnce | 4847,67 | [4828.75; 4866.58] |
| numpyIntegerPhiloxAtOnce | 4849,22 | [4829.80; 4868.63] |
| tensorflowIntegerAtOnce | 4893,72 | [4862.41; 4925.04] |
| pythonIntegerOneByOne | 5410,52 | [5185.95; 5635.09] |
| numpyIntegerOneByOne | 5925,94 | [5579.41; 6272.47] |
| pytorchIntegerOneByOne | 8846,44 | [8492.64; 9200.24] |
| numpyIntegerMtOneByOne | 45223,56 | [42623.11; 47824.02] |
| numpyIntegerPhiloxOneByOne | 64212,81 | [61111.12; 67314.50] |

Table 5: Energy consumption in Joule by minutes for each experiments with 32 bits integer

| Generator | Energy consumption (J/min) | Energy consumption 95% CI |
|---|---|---|
| mrg32k3aO2 | 2750,80 | [2744,56; 2757,03] |
| mrg32k3a | 2783,59 | [2744,86; 2822,31] |
| well19937O2 | 2979,42 | [2970,81; 2988,02] |
| well19937O3 | 2991,04 | [2981,97; 3000,10] |
| mrg32k3aO3 | 3000,87 | [2941,58; 3060,15] |
| well19937a | 3040,66 | [3008,42; 3072,90] |
| mt19937arO2 | 3179,17 | [3168,29; 3190,04] |
| mt19937ar | 3186,49 | [3172,36; 3200,62] |
| mt19937arO3 | 3226,83 | [3159,98; 3293,67] |
| pythonOneByOne | 3865,71 | [3806,83; 3924,59] |
| pcg64O3 | 3882,55 | [3812,37; 3952,73] |
| pcg64O2 | 3925,18 | [3872,56; 3977,79] |
| pythonTasksetAtOnce | 3994,90 | [3871,83; 4117,96] |
| pytorchTasksetAtOnce | 4348,02 | [4306,74; 4389,30] |
| numpyMtAtOnce | 4358,07 | [4168,48; 4547,65] |
| pcg64 | 4473,07 | [4459,61; 4486,52] |
| numpyTasksetAtOnce | 4550,28 | [4530,53; 4570,03] |
| tensorflowTasksetAtOnce | 4762,63 | [4750,04; 4775,22] |
| numpyPhiloxAtOnce | 5033,01 | [4875,79; 5190,22] |
| tensorflowAtOnce | 5131,12 | [5100,87; 5161,37] |
| pytorchOneByOne | 5630,90 | [5233,18; 6028,62] |
| numpyAtOnce | 5906,65 | [5670,16; 6143,13] |
| numpyOneByOne | 5935,29 | [5772,42; 6098,16] |
| pythonAtOnce | 6521,71 | [6078,98; 6964,44] |
| pytorchAtOnce | 7141,50 | [6891,12; 7391,88] |
| numpyMtOneByOne | 37375,55 | [36671,35; 38079,75] |
| numpyPhiloxOneByOne | 37590,44 | [36613,97; 38566,92] |

Table 6: Energy consumption in Joule by minutes for each experiments, on 64 bits double.

## 4.3 Overall energy consumption

Tables 7 and 8 illustrate the energy consumption in Joule during the real time taken by each experiment (e.g. $2^{30}$ random number generation, depending on the algorithm and the technology). Last tables were about the energy consumption in Joule by minutes, but we are looking here at the energy consumption during the time taken for each experiment. While C implementations are more energy-efficient per minute, the competitive overall execution time of Python enables it to rival C implementations. Underlying implementations of machine learning technologies in Python are often using C, C++ or CUDA. For 32-bit integers, the PCG algorithm demonstrates notable efficiency, outpacing other C implementations and followed closely by NumPy, while also maintaining reasonable energy consumption given its execution time. The Mersenne Twister algorithm in C exhibits the highest consistency, yielding similar results in both integer and double generation. Regrettably, aside from MT and PCG, other C-based PRNGs are dedicated to a specific output type. For instance Mrg32k3a and Well (Panneton et al., 2006) are dedicated to generating double values and Philox is generating integers. Unlike Mersenne Twister, PCG displays a significant discrepancy between its integer and double generation performance. Although O2 or O3 optimization does not affect per-minute energy consumption, its reduction of total computation time contributes to lower the overall energy needed. It is noteworthy that for PCG and Philox (in 32-bits generation), O2 and O3 optimizations were not applied since the use of these optimizations causes code malfunctions, resulting in immediate termination without executing the intended operations. The O3 optimization level is known as aggressive and may often produce non reproducible results or strange behavior and this is documented. However it is the first time in more than 30 years of computer experiments that we observe the O2 level producing unusable and thus non reproducible results.

| Generator | Energy consumption during real time (J) | Energy consumption during real time 95% CI |
|---|---|---|
| pcg32Integer | 131,17 | [130.18; 132.17] |
| numpyIntegerTaskseAtOnce | 206,41 | [205.60; 207.22] |
| mt19937arIntegerO3 | 258,09 | [248.73; 267.46] |
| tensorflowIntegerAtOnce | 262,52 | [260.84; 264.20] |
| mt19937arIntegerO2 | 262,77 | [255.07; 270.48] |
| numpyIntegerAtOnce | 267,33 | [266.28; 268.38] |
| numpyIntegerMtAtOnce | 367,77 | [366.33; 369.20] |
| mt19937arInteger | 404,65 | [388.90; 420.41] |
| numpyIntegerPhiloxAtOnce | 547,38 | [545.19; 549.58] |
| tensorflowIntegerTasksetAtOnce | 566,63 | [564.73; 568.53] |
| pytorchIntegerTasksetAtOnce | 640,54 | [638.00; 643.08] |
| pytorchIntegerAtOnce | 728,84 | [723.20; 734.47] |
| philoxInteger | 7273,10 | [7241.08; 7305.12] |
| pythonIntegerTasksetAtOnce | 34827,75 | [34514.05; 35141.44] |
| pythonIntegerAtOnce | 35659,70 | [33502.49; 37816.92] |
| pythonIntegerOneByOne | 38407,42 | [36813.25; 40001.59] |
| pytorchIntegerOneByOne | 336428,77 | [322973.74; 349883.80] |
| numpyIntegerOneByOne | 637889,20 | [600587.67; 675190.72] |
| numpyIntegerMtOneByOne | 4769399,46 | [4495148.41; 5043650.52] |
| numpyIntegerPhiloxOneByOne | 7012261,80 | [6673546.74; 7350976.87] |

Table 7: Energy consumption in Joule during real computation time, for 32 bits integers.

| Generator | Energy consumption during real time (J) | Energy consumption during real time 95% CI |
|---|---|---|
| mt19937arO3 | 225,74 | [221,06; 230,42] |
| mt19937arO2 | 238,60 | [237,79; 239,42] |
| well19937O2 | 247,04 | [246,33; 247,75] |
| well19937O3 | 247,41 | [246,66; 248,16] |
| tensorflowAtOnce | 288,69 | [286,99; 290,40] |
| numpyTasksetAtOnce | 330,27 | [328,83; 331,70] |
| mt19937ar | 397,49 | [395,73; 399,26] |
| pytorchTasksetAtOnce | 436,27 | [432,13; 440,41] |
| numpyAtOnce | 568,32 | [545,56; 591,07] |
| tensorflowAtOnce | 649,36 | [647,65; 651,08] |
| well19937a | 663,01 | [655,98; 670,04] |
| pcg64O3 | 711,70 | [698,83; 724,56] |
| pcg64O2 | 724,40 | [714,69; 734,11] |
| pytorchAtOnce | 821,43 | [792,63; 850,23] |
| numpyMtAtOnce | 950,24 | [908,91; 991,58] |
| pcg64 | 982,61 | [979,65; 985,56] |
| mrg32k3aO3 | 998,58 | [978,85; 1018,31] |
| numpyPhiloxAtOnce | 1111,94 | [1077,21; 1146,68] |
| mrg32k3aO2 | 1442,61 | [1439,33; 1445,88] |
| mrg32k3a | 2000,77 | [1972,93; 2028,61] |
| pythonOneByOne | 2375,43 | [2339,25; 2411,61] |
| pythonTasksetAtOnce | 4629,02 | [4486,43; 4771,62] |
| pythonAtOnce | 8202,27 | [7645,45; 8759,08] |
| numpyOneByOne | 32740,55 | [31842,13; 33638,98] |
| numpyMtOneByOne | 199270,14 | [195515,66; 203024,62] |
| numpyPhiloxOneByOne | 202400,80 | [197143,08; 207658,52] |
| pytorchOneByOne | 224148,23 | [208316,06; 239980,39] |

Table 8: Energy consumption in Joule during real computation time, for 64 bits double.

## 4.4   Statistical quality

Now, we examine the quality of the pseudo random numbers generated by each technology. Integer results are presented in table 9, and double results are presented in table 10. First, we notice that the quality of the double generation behave more as expected than integers. Indeed, each PRNG algorithm is known to fail specific Big Crush tests, so we can use the tests as markers to recognize one PRNG or another. In the double generation, all implementations of the Mersenne Twister algorithm—including MT in C, Python, and NumPy—failed to the LinearComp tests 80 and 81, aligning with expectations since those tests are linked to crypto-security. The Well algorithm also demonstrated similar failures, its internal structure is similar to MT with huge feedback shift registers. Conversely, PCG and its NumPy variant employing the PCG 128/64 XSL-RR algorithm passed all tests, corroborating the assertions of the author of PCG. The Philox algorithm from NumPy failed the BirthdaySpacings test, in contrast to its TensorFlow counterpart, which passed all assessments. From our experience, we acknowledge that PRNGs may occasionally fail tests they are not expected to (Antunes & Hill, 2023). Further scrutiny into each behavior of PRNG would necessitate multiple replications with varying initial statuses to verify consistency across the entire period. Regrettably, we had to exclude PyTorch data from table 10 due to its failure in 62 statistical tests (among 106 tests). This result require additional investigation, this high failure rate suggests inferior statistical quality (58% of the big crush battery failed). Interestingly, the pseudo random number generation of PyTorch on 32-bit integers was way better, failing only 3 tests.

Concerning the 32 bits integers, results were a bit more surprising. First, we can notice that, while the original C code Mersenne Twister fails the two tests 80 and 81, the NumPy MT implementation only failed one test. Surprisingly, the Python version passed all tests, more investigations with different initial statuses could be interesting. PCG and Philox, in their different implementations, did fail some tests, but still remains good quality generators. It is interesting to note that they do indeed fails some tests, while authors assume that they do not fail any. In addition, they did not fail the same tests. For example, NumPy version of Philox failed the test 49 MaxOft, while the TensorFlow version of Philox failed the test 9 CollisionOver. This makes us think that these PRNGs might be failing different statistical tests if we would try to do replications over the PRNGs periods, using different state of the PRNG. An unexpected observation was the failure of PyTorch at three specific tests, notably the RandomWalk and two LinearComp tests. Although the documentation of PyTorch suggests Philox as its underlying PRNG, the observed failures look like the "signature" of a Mersenne Twister, raising questions about its implementation.

| Generator | Number of Failed Tests | Failed Tests |
|---|---|---|
| philoxInt32 | 6 | 34 Gap, 35 Gap, 36 Gap, 37 Gap, 65 SumCollector, 68 MatrixRank |
| pytorchInt32 | 3 | 77 RandomWalk1, 80 LinearComp, 81 LinearComp |
| pcgInt32 | 1 | 5 CollisionOver |
| mtInt32 | 2 | 80 LinearComp, 81 LinearComp |
| numpyMtInt32 | 1 | 80 LinearComp |
| numpyPhiloxInt32 | 1 | 49 MaxOft |
| numpyInt32 | 0 | |
| tensorflowInt32 | 1 | 9 CollisionOver |
| pythonInt32 | 0 | |

Table 9: Failed BigCrush tests for each experiments, based on 32 bits integer.

| Generator | Number of Failed Tests | Failed Tests |
|---|---|---|
| pcgReal | 0 | |
| tensorflowReal | 0 | |
| MRG32k3aReal | 0 | |
| wellReal | 2 | 80 LinearComp, 81 LinearComp |
| numpyReal | 0 | |
| mtReal | 2 | 80 LinearComp, 81 LinearComp |
| numpyPhiloxReal | 1 | 21 BirthdaySpacings |
| pythonReal | 2 | 80 LinearComp, 81 LinearComp |
| numpyMtReal | 2 | 80 LinearComp, 81 LinearComp |

Table 10: Failed BigCrush tests for each experiments, based on 64 bits double. PyTorch excluded for readability, failing 62 tests.

## 4.5 Numerical reproducibility

An important finding of this study is the absence of reproducibility in the numbers generated across various platforms. The Mersenne Twister algorithm, initialized with the same seed, will give different numbers with the different C, Python and NumPy implementations. The same applies for PCG with NumPy and C code, and also for Philox with NumPy and TensorFlow frameworks. The reason might be because we are using seeds to initialize our PRNG, instead of setting the full state. This function transforming a seed into a full state might differ between the technologies or frameworks, and this can lead to a loss of numerical reproducibility between the technologies.

## 5 DISCUSSIONS

From what has been discovered comes more questions: if the loss of reproducibility does not come from the seeding functions, this leads us to a critical inquiry: how can we ascertain that the algorithm in use is a correct implementation of the generator? For us, this is an open question. The ideal solution would be for original authors to supply a sample of generated pseudo-random numbers, which we should be able to compare with the numbers we are generating, to ensure perfect reproducibility. To our knowledge, Mersenne Twister is the only PRNG that offers this feature with the expected output. Numerical reproducibility is not only important for the advancement of Science but also for debugging (Hill, 2015). Does the change of hardware or software stack affect the reproducibility of a PRNG? What we observe is that the portability of PRNGs should not be considered as granted. Here, we are using different technologies with the same environment, and we obtained different results and different statistical quality. Another way to identify the PRNG not based on the numerical result would be to perform statistical tests to try to identify the underlying algorithm, as some failed statistical tests can serve as markers for some PRNGs. However, as we found in a deep study (Antunes & Hill, 2023), the same PRNG algorithm might fail several different statistical tests. For example, over 4096 replications, it appears that the Mersenne Twister algorithm fails the all 106 BigCrush tests at least once. We could expect a similar behavior from other PRNGs. This would also need further investigations. Ensuring the use of a specific algorithm, in the absence of perfect numerical reproducibility, is far from trivial.

In the discussions surrounding high-performance computing (HPC), it is undeniable that it is a high-consuming endeavor in terms of time, financial investment, and energy. With the inexorable march towards greater computational power and despite technological innovation, these costs have only intensified due to inflation in hardware, energy prices and also the Jevons paradox. Meanwhile, ML has emerged as an indispensable tool in a plethora of fields such as the now famous Large Language Models (LLMs), but also in more common domains like autonomous vehicles, healthcare, and so on. The sophistication of ML models comes with its own demands on computational and energy resources. When considering the generation of pseudo-random numbers — an essential component for stochastic processes, simulations, and even for the operation of ML algorithms themselves — the comparison between traditional C-coded generators such as Philox, Mersenne Twister, and PCG, and those implemented within ML frameworks (using PyTorch, TensorFlow, Python, and NumPy), presents a complex picture. Energy consumption is a critical factor; while there is no actual data on the exact energy costs of random number generation within neural network training, it is reasonable to assume that the proportion is non-negligible. Profiling such applications to evaluate the exact proportion of time used in the generation of pseudo random numbers, depending on the size of the neural network, would be valuable. With a neural network like GPT-4 LLM provided by OpenAI, we have around 175 billion of parameters (edges of the graph), we can easily imagine that a very large number of pseudo random numbers have been used. Generating pseudo-random numbers is an integral part of the training phase of neural networks, especially in processes such as weight initialization, shuffling, and during stochastic gradient descent where randomness is used to ensure convergence. The results of our investigation suggest that ML implementations can match the statistical quality and speed of their C code counterparts. However, the ease and speed of generating pseudo-random numbers using ML frameworks leads to a little increase of the energy consumption cost dedicated to this task (around 10%).

In future works, we want to ascertain that Powerjoular, utilizing RAPL, provides reliable measurements of energy consumption. In (Khan et al., 2018), they tested the reliability of RAPL, on a Finnish supercomputing cluster, and on Amazon EC2, leading to the conclusion that RAPL is accurate and have negligible performance overhead. Powerjoular adds a layer on top of this. We reasonably think that our results are reliable but we want to check it more thoroughly. Another limitation, which also presents a research opportunity, is the measurement of the impact of the PRNG quality on neural network training.

## CONCLUSION

Machine learning frameworks use Pseudo-Random Number Generators (PRNGs) in the training of neural networks. The input of stochastic sources has proven to be interesting for the machine learning field. Yet, the investigation into the quality of generated pseudorandom numbers, along with their generation time and power requirements, remains incomplete. This study evaluated the efficiency of pseudo-random number generation within machine learning technologies as compared to traditional C language implementations. We examined Python, PyTorch, TensorFlow, and NumPy. Our findings indicate that the different libraries and frameworks using the Python language are well-optimized. The NumPy library stands out in terms of time consumption and quality, aligning closely with C implemented PRNGs. Nevertheless, two drawbacks were identified: firstly, machine learning technologies tend to consume approximately 10% more energy; secondly, there is an inconsistency in numerical repeatability when using identical seeds with different PRNGs implementations, this is a portability problem. This brings into question the fidelity to the initial PRNG implementations according to their original descriptions. The implementation of PRNGs in machine learning tools should give identical results when initialized similarly to their C counterparts. Despite claims on the official PCG website labeling it as "Very fast" compared to the "Acceptable" speed of Mersenne Twister, our analysis suggests that such claims are exaggerated and even false since the generation of double values essential for so many simulations is 2.5 times faster with the original MT (we noticed some differences of performance between the generation of 32 bits integer and 64 bits double pseudo random values). The C implementation of PCG, takes the same time than the NumPy implementation. In addition, PCG also exhibited failures in certain BigCrush tests though it was claimed crush resistant. PCG should be avoided for massive parallel computing as stated in the NumPy documentation which suggests to use PCG64DXSM. Additional research is needed for a deeper exploration for each PRNG. Although the impact of PRNG quality on neural network training outcomes has not been extensively researched, insights from the following recent studies discussed in section 2 suggest that PRNG quality could indeed influence the performance of the resulting neural network, based on quality metrics (Huk et al., 2021; Koivu et al., 2022).

## CREDIT AUTHORSHIP CONTRIBUTION STATEMENT

**Benjamin Antunes**: Writing–original draft, Conceptualization, Methodology, Software, Formal analysis, Investigation, Data curation.
**David R.C. Hill**: Supervision, Writing – review & editing, Validation.

## DECLARATION OF COMPETING INTEREST

The authors declare that they have no known competing financial interests or personal relationships that could have appeared to influence the work reported in this paper.

## AUTHOR BIOGRAPHIES

**BENJAMIN ANTUNES** is a Phd Student at Clermont Auvergne University (UCA). He holds a Master in Computer Science (head of the list). His thesis subject is about the reproducibility of numerical results in the context of high performance computing. He is espacially working on reproducibility issues in parallel stochastic computing. His email address is benjamin.antunes@uca.fr and his homepage is https://perso.isima.fr/~beantunes/

**DAVID HILL** is doing his research at the French Centre for National Research (CNRS) in the LIMOS laboratory (UMR 6158). He earned his Ph.D. in 1993 and Research Director Habilitation in 2000 both from Blaise Pascal University and later became Vice President of this University (2008-2012). He is also past director of a French Regional Computing Center (CRRI) (2008-2010) and was appointed two times deputy director of the ISIMA Engineering Institute of Computer Science – part of Clermont Auvergne INP, #1 Technology Hub in Central France (2005-2007 ; 2018-2021). He is now Director of an international graduate track at Clermont Auvergne INP. Prof Hill has authored or co-authored more than 250 papers and has also published several scientific books. He recently supervised research at CERN in High Performance Computing. ( https://isima.fr/~hill/  -  david.hill@uca.fr )